\documentclass[aps,prl,twocolumn]{revtex4}
\usepackage{amsmath}
\usepackage{graphicx}
\usepackage[section]{placeins}

\newcommand{\ket}[1]{|#1\rangle}
\newcommand{\bra}[1]{\langle#1|}

\newcommand{\eq}{\begin{equation}}
\newcommand{\fine}{\end{equation}}

\begin{document}

\title{Experimental entanglement restoration on entanglement-breaking channels}
\author{Fabio Sciarrino$^{2,1}$, Eleonora Nagali$^{1}$, Francesco De Martini$^{1,3}$%
, Miroslav Gavenda$^{4}$, and Radim Filip$^{4}$ \\
$^{1}$Dipartimento di Fisica dell'Universit\'{a} ''La Sapienza'' and
Consorzio Nazionale Interuniversitario per le Scienze Fisiche della Materia,
Roma 00185, Italy\\
$^{2}$Centro di\ Studi e Ricerche ''Enrico Fermi'', Via Panisperna
89/A,Compendio del Viminale, Roma 00184, Italy\\
$^{3}$Accademia Nazionale dei Lincei, via della Lungara 10, Roma 00165, Italy\\
$^{4}$Department of Optics, Palack$\acute{y}$ University, 17. Listopadu 50, Olomuc 77200, Czech Republic}

\begin{abstract}
\end{abstract}

\maketitle

\textbf{Quantum entanglement, a fundamental property ensuring security of
key distribution and efficiency of quantum computing, is extremely
sensitive to decoherence. Different procedures have been developed in order
to recover entanglement after propagation over a noisy channel. However,
besides a certain amount of noise, entanglement is completely lost. In this
case the channel is called entanglement breaking and any multi-copy
distillation methods cannot help to restore even a bit of entanglement.
We report the experimental realization of a new method which restores
entanglement from a single photon entanglement breaking channel. The method
based on measurement of environmental light and quantum feed-forward correction
can reveal entanglement even if this one completely disappeared. This protocol provides new elements to overcome decoherence effects.}%
\newline
The resource of quantum entanglement lies at the basis of many protocols of
quantum communication and information \cite{Horo07}. In order to face the problems
introduced by the real world, it is primary of importance to consider the
unavoidable interaction between noise and signal. The presence of noise
alters or even invalidates the transmission of quantum information through
communication channel, by spoiling entanglement. Nowadays many
investigations have been developed in order to restore the entanglement
lost, performing purification and distillation protocols \cite{Pan03}. However, all these
techniques can operate only on systems characterized by non-vanishing entanglement. When a channel is
entanglement-breaking \cite{Horo03}, that is, no entanglement is left after the
interaction with noise, it is insufficient to work only at the input and
output of the channel, and one has to go into the channel process and try to
get more information to reveal entanglement. Thus a physical structure of
the channel starts to be relevant. An interesting approach is based on an
extension of \textit{quantum erasing} idea \cite{Scul89}, \cite{Engl95}, \cite{Zajo91}, \cite{Kwia92}: a state of the system can be restored
if a proper measurement is performed on the environment \cite{Greg03}. This procedure,
recently introduced as environmental channel correction \cite{Busc05}, \cite{Hayd05}, \cite{Smol05}, offers a
theoretical possibility to deterministically restore complete entanglement. This is an interesting perspective, but the
environment is normally not under complete control. Naturally, it is
desirable to recover entanglement only by measuring that part of the environment
which is directly and closely coupled to the signal and to relax any
constraints on the environment previous to interaction and on the noise
coupling. The protocol introduced here is able to restore entanglement
without control on the environment previously to the interaction. Since no information is directly
communicated by the entangled pair, we can reasonably relax from the
deterministic correction and assume rather general probabilistic
corrections, similarly as for distillation protocol.\newline

\begin{figure}[t]
\centering
\includegraphics[scale=.35]{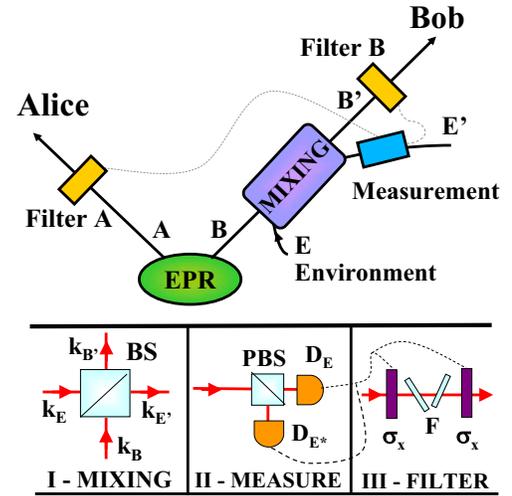}
\caption{Schematic representation of the entanglement restoration procedure. 
Generation by the EPR source of the entangled photons shared
between Alice and Bob. On Bob mode is showed the interaction between the
entangled photon and the environment and the measure on the environmental
photon that drives the filtration process.}
\end{figure}

The overall dynamic of the protocol can be divided in three different steps:\newline
\textbf{I) Coupling with environment.} Alice (A) and Bob (B) share a
polarization entangled state $|\Psi ^{-}\rangle _{AB}=\frac{(\ket{H}_{A}\ket{V}_{B}-i\ket{V}_{A}\ket{H}_{B})}{\sqrt{2}}$, and the photon $B$
propagates over a noisy channel. The environment $E$ is
represented by a completely unpolarized photon described by the density
matrix ${\rho}_{E}=\frac{{I}}{2}=\frac{\ket{H}\bra{H}+\ket{V}\bra{V}}{2}$ that interacts with $B$ by
localizable simple linear coupling: a generic beam splitter BS with
transmittivity $T$ and reflectivity $R=1-T$. All the results that will be
presented can be directly extended for general passive coupling between the
two modes. After the interaction on the beam splitter, three possible
situations can be observed: both the photons go to environment or to the
signal mode, or only a single photon is separately presented in signal and
environment. The first case corresponds simply to attenuation, while the
second case can be in principle distinguished by counting the number of
photons in the signal. Thus only the last case is interesting to be
analyzed. After the interaction process, the entanglement has been damaged
and, below a threshold value of $T$, no entanglement can be observed in the
output state.

\textbf{II)Measurement of the environment.} We present here a procedure
which restores the entanglement by measuring the photon on the environmental
mode. The photon propagating on mode $k_{E^{\prime }}$ is measured after a
polarization analysis realized through a polarizing beam splitter (PBS)(Fig.1-\textbf{II}). Two different scenarios will be analyzed: in the first
one, the environmental photon and the signal one are \textit{distinguishable}
in principle but, due to technological limitations, we are not able to
achieve a discrimination among them. In this situation the restoration is able
to partially recover entanglement. On the other hand, when the two photons
are \textit{indistinguishable}, the protocol exploits quantum interference
phenomena, analogously to the quantum state teleportation, and
asymptotically retrieves all the initial entanglement.

\begin{figure}[t]
\centering
\includegraphics[width=0.5\textwidth]{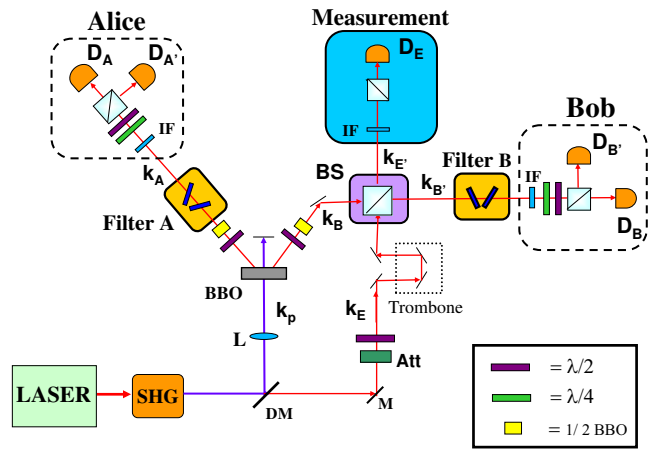}
\caption{Experimental setup. The main source of the experiment is a Ti:Sa
mode-locked laser with wavelength (wl) $\protect\lambda =795nm$. A
small portion of this laser,
generates the single noise photon over the mode $k_{E}$ using an attenuator ($ATT$%
). The transformation used to map the state $\left| H\right\rangle _{E}$
into $\protect\rho _{E}=\frac{I_{E}}{2}$ is achieved either adopting a
Pockels cell driven by a sinusoidal signal, either through a stochastically
rotated $\protect\lambda /2$ waveplate \protect\cite{Scia04} . The main part of the laser
through a second harmonic generation (SHG)which generates a UV laser beam with wave-vector
$k_{p}$ and power equal to $800mW$. This field pump a $1.5mm$ thick non-linear crystal of $\protect\beta $%
-barium borate (BBO) cut for type II phase-matching that generates
polarization entangled pairs $\left| \Psi ^{-}\right\rangle _{AB}$ with
equal wavelength $\protect\lambda =795nm$. The spatial and temporal walk-off
is compensated through a $\frac{\protect\lambda }{2}$ waveplate and a$\
0.75$ mm thick BBO \cite{Kwia95}. The
dashed boxes indicate the polarization analysis setup adopted by Alice and
Bob. The photons are coupled to a single mode fiber and detected by single
photon counting modules $D_{i}$. On output modes $k_{B^{\prime }}$ and 
$k_{E^{\prime }}$ the photons are spectrally filtered adopting two
interference filters (IF) with bandwidth equal to $3nm$, while on mode $%
k_{A}$ the IF has bandwidth $4.5nm$. For indistinguishable noise photon the IF on $k_{B^{\prime }}$ and $k_{A}$ have been replaced by IF with FWHM equal to $1.5nm$. The output signals of the
detectors are sent to a coincidence box interfaced with a computer, which
collects the different double and triple coincidence rates.
The detection of triple coincidence ensures the presence of one photon per
mode.}
\label{fig:Mth_order}
\end{figure}


\begin{table*}[t!]
\begin{center}
\textbf{(a) Noise distinguishable photons}\\
\footnotesize
\begin{tabular}{||c||c|c||c|c|c|c|c||c|c||}
\hline \hline
$\rho_{AB}$ & $A_{A}$ & $A_{B}$ & $\alpha$ & $\beta$ & $\gamma$ & $\delta$ & $\xi$ & $C$  & $P$    \\
\hline \hline
$I$ &  $ $  & $ $ & $R^2$ & $R^2 + 2T^2$ & $R^2 + 2T^2$ & $R^2$  & $-2T^2$ & $Max(0,\frac{2T^2-R^2}{2(R^2+T^2)})$ & $R^2+T^2$ \\
\hline
$II$ & $ $  & $ $ & $0$ & $T^2$ & $R^2 + T^2$ & $R^2$  & $-T^2$ & $\frac{T^2}{T^2+R^2}$ & $\frac{1}{2}(R^2+T^2)$ \\
\hline
$III$ & $\varepsilon \frac{T^2}{T^2+R^2}$ & $\varepsilon $ & $0$ & $\varepsilon T^2$ & $\varepsilon T^2$ & $\varepsilon^2 R^2$ & $-\varepsilon \frac{T^3}{\sqrt{T^2+R^2}}$ & $\frac{1}{1 + \varepsilon \frac{R^2}{2 T^2}}\frac{T}{\sqrt{T^2+R^2}}$ & $\frac{1}{2} \varepsilon T^{2} + \frac{1}{4} {\varepsilon}^2 R^2$ \\
\hline \hline
\end{tabular}\\
\vspace{0.2cm}
\small
\textbf{(b) Noise indistinguishable photons}\
\footnotesize{
\scriptsize
\begin{tabular}{||c||c|c||c|c|c|c|c||c|c||}
\hline \hline
$\sigma_{AB}$ & $A_{A}$  & $A_{B}$ & $\alpha$ & $\beta$ & $\gamma$ & $\delta$ & $\xi$ &  $C$  & $P$    \\
\hline \hline
$I$ & $ $ & $ $ & $R^2$ & $1-4T+5T^2$ & $R^2+2T^2-2RT$ & $R^2$ & $-2T^2+2RT$ & $Max(0,\frac{2T\left|T-R\right|-R^2}{2P})$ & $R^2+T^2-RT$ \\
\hline
$II$ &  $ $ & $ $ & $0$ & $T^2$ & $(2T-1)^2$ & $R^2$  & $-T^2+RT$ & $\frac{T\left|T-R\right|}{2P}$ & $\frac{1}{2}(R^2+T^2-RT)$ \\
\hline

$\begin{tabular}{c}$III$\\ $\footnotesize T>\left|2T-1\right|$\end{tabular}$ & $\varepsilon $ & $\varepsilon \left(\frac{2T-1}{T}\right)^2 $ & $0$ & $\footnotesize \varepsilon T^2$ & $\footnotesize\varepsilon T^2$ & $\scriptsize\varepsilon^2 \left(\frac{TR}{2T-1}\right)^2$  & $-\varepsilon T^2$  &  $\scriptsize\frac{1}{1+\frac{\varepsilon R^2}{2(2T-1)^2}}$ & $\frac{1}{4}(\scriptsize 2 \varepsilon T^2 + \varepsilon^2 \left(\frac{TR}{2T-1}\right)^2)$ \\
\hline
$\begin{tabular}{c}$III$\\ $T<\left|2T-1\right|$\end{tabular}$ & $\footnotesize\varepsilon \scriptsize \left(\frac{T}{2T-1}\right)^2$ & $\footnotesize \varepsilon $ & $\footnotesize 0$ & $\varepsilon (2T-1)^2$ & $\varepsilon (2T-1)^2$ & $\footnotesize \varepsilon^2 R^2$  & $-\varepsilon (2T-1)^2$ & $\frac{1}{1+\frac{\varepsilon R^2}{2(2T-1)^2}}$ & $\frac{1}{4}(2\varepsilon (2T-1)^2 + \varepsilon^2 R^2)$ \\
\hline \hline\end{tabular}}
\end{center}
\noindent TABLE I. Theoretical values of the elements of ${\rho}_{AB}$ matrix, concurrence C, probability of success P, and intensity $A$ of filtering on Alice and Bob mode, for the three different steps of the restoration procedure.
\normalsize
\end{table*}


\textbf{III)Filtration.} A key role in the protocol is represented by a
filtration performed both on Alice and Bob modes. Such operation leads to a higher entanglement of the bipartite system, at the cost of a lower probability of
implementation. As shown in Fig.(1-\textbf{III}), the filtering is achieved by two
sets of glass positioned close to their Brewster's angle, in order to
attenuate one polarization ($V$) in comparison to its orthogonal ($H$). We indicate the
attenuation over the mode $k_{i}$ for the $V$ polarization as $A_{i}$: $\ket{V}_{i}\rightarrow \sqrt{A_{i}}\ket{V}_{i}$. By tuning the incidence angle of the beam on mode $k_{i}$, different values of
attenuations ${A_{i}}$ can be achieved. The complete protocol
implies a classical feed-forward on the polarization state of the photon
belonging to the mode $k_{B^{\prime }}$ depending on which detector on environmental mode ($D_{E}$
or $D_{E}^{\ast }$, Fig.1-\textbf{II}) fires: precisely if the detector $D_{E}$ clicks no
transformation is implemented on the quantum channel, in the other case two $%
\sigma _{X}$ are applied before and after the filtration: Fig.(1-\textbf{III}). This
conditional operation could be realized adopting the electronic scheme
experimentally demonstrated in \cite{Giac02,Scia06}. Without feedforward,
the efficiency of the overall procedure is reduced by a factor 2.%
\newline

The different stages of the restoration process will be described by the
theoretical density matrices ${\rho}_{AB}$, written in the basis $%
\{|HH\rangle ,|HV\rangle ,|VH\rangle ,|VV\rangle ,\}$ as:
\begin{equation*}
{\rho}_{AB}=\frac{1}{4P}\left( 
\begin{array}{cccc}
\alpha  & 0 & 0 & 0 \\ 
0 & \beta  & i\xi  & 0 \\ 
0 & -i\xi  & \gamma  & 0 \\ 
0 & 0 & 0 & \delta 
\end{array}
\right) 
\end{equation*}
where the parameters $\alpha ,\beta ,\gamma ,\delta ,\xi $ vary in the
different protocol steps and $P$ represents a normalization parameter related
to the probability of each operation. The degree of entanglement after each step is evaluated through the concurrence C, which assumes values $0\leq C \leq 1$ dependently if the state is separable ($C=0$) or entangled ($C >0$). In particular, a maximally entangled state reaches $C=1$ \cite{Woot98}.

A schematic drawing of our experimental layout is shown in Fig.2.
In order to couple the noise with the signal, the photons on modes $k_{B}$
and $k_{E}$ are injected in the two input arms of a beam-splitter ($BS$) 
with transmittivity $T$. A mutual delay $%
\Delta t$, micro-metrically adjustable by a two-mirror optical ''trombone'',
can change the temporal matching between the two photons. The
setting value $\Delta t=0$ corresponds to the full overlapping of the photon
pulses injected into $BS$, i.e., to the maximum photon interference.

\textbf{Distinguishable noise photons}\newline
The experiment with distinguishable photons has been achieved by injecting a
single photon $E$ with a mutual delay with photon $B$ of $\Delta t>>\tau
_{coh}=300 fs.$ We note that the resolution time of the detector $t_{\det
}>>\Delta t,$ hence it is not technologically possible to individuate
whether the detected photon belongs to the environment or to the entangled
pair. 

\textbf{I)} Without any access to the environmental photon, after the mixing
at the beam splitter the state $|\Psi^{-}\rangle_{AB}$ evolves into
the density matrix ${\rho}^{I}_{AB}$ (see Table.I). As can be observed in Fig.(3-\textbf{a}), for $T\leq (\sqrt{2}-1)$ the concurrence $C_{I}$ vanishes.
To verify the theory we carried an experiment by adopting a beam splitter with $T\approx0.4$.
The experimental density matrix $\widetilde{\rho }_{AB}^{I}$ is shown in
Fig.(4-\textbf{I}) and has a high fidelity with the theoretical prediction: $F_{I}=F(\widetilde{\rho}_{AB}^{I},{\rho}_{AB}^{I})=(0.997\pm 0.006)$, where $F(\rho ,\sigma)=Tr^{2}%
\left[ \left( \sqrt{\rho }\sigma \sqrt{\rho }\right) ^{1/2}\right]$ represents the fidelity between two mixed states $\rho$ and $\sigma$. As expected, the state obtained experimentally exhibits $\widetilde{C}_{I}=0$, hence the communication channel is entanglement breaking.

\begin{figure}[h]
\centering
\includegraphics[width=0.5\textwidth]{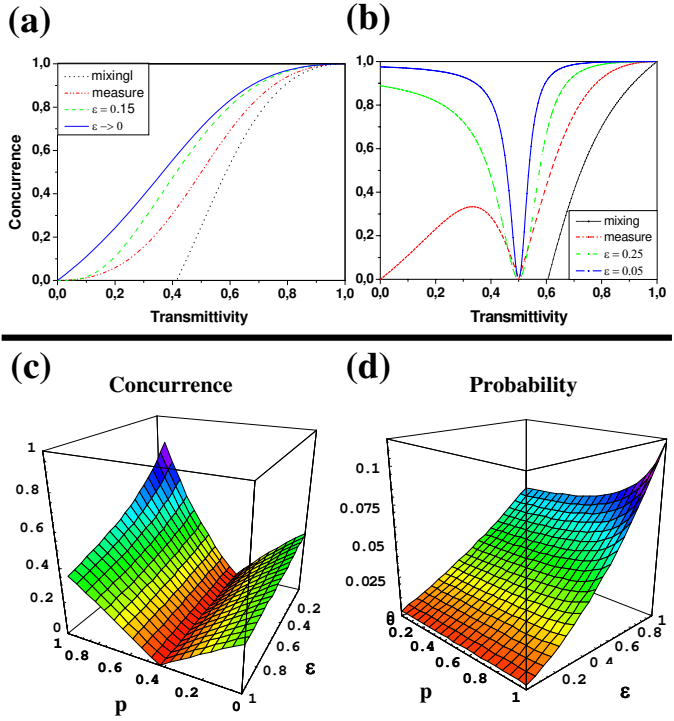}
\caption{(\textbf{a-b}) Concurrence for different scenarios \textbf{a})
Distinguishable photons - without measurement (dotted line), with
measurement (dashed -dotted line), with measurement and  filtration $%
\protect\varepsilon =0.15$ (dashed line), with measurement and 
filtration $\protect\varepsilon \rightarrow 0$ (continuous line). \textbf{b}%
) Indistinguishable photons - without measurement (dotted line), with
measurement (dashed -dotted line), with measurement and filtration $%
\protect\varepsilon =0.25$ (dashed line), with measurement and
filtration $\protect\varepsilon =0.05$ (continuous line). (\textbf{c-d})
Strategy with measurement and filtration applied to a mixture of
indistinguishable and distinguishable photons where $p$ quantifies the
degree of indistinguishability ($T=0.3$): \textbf{c}) concurrence$,$ \textbf{d})
Probability of implementation.}
\end{figure}

\textbf{II)} As no selection has been performed on the environment before the coupling with signal, a multimode noise interacts with the entangled photon. After the interaction, a single mode fiber has been inserted on the output mode of the BS in order to select only the output mode connected with the entanglement breaking. Let us consider the case in which the photon on mode $k_{E}$ is measured in the state $|H\rangle_{E}$ (Fig.1-\textbf{II}). The state evolves into an entangled one described by the density
matrix ${\rho}^{II}_{AB}$. Theoretically, the
entanglement is restored for all the values of $T\neq0,\frac{1}{2}$ but the elements of the
matrix are fairly unbalanced. The experimental result $\widetilde{\rho }%
_{AB}^{II} $ is reported in Fig.(4-\textbf{II}). According to theory, we expect a restoration of the entanglement with $C_{II}=0.32$, and success rate $P_{II}=0.27$. The experimental state $\widetilde{\rho}_{AB}^{II}$ is characterized by a fidelity with the theoretical one $F_{II}=(0.96 \pm 0.06)$, which leads to $\widetilde{C}_{II}=(0.19\pm 0.02)>0$ and $\widetilde{P}_{II}=(0.26\pm 0.01)$. 

\textbf{III)} If T is known then the state can be further locally
filter out. Two  filters $F_{A}$ and $F_{B}$ acting on the modes $k_{A}$ and $k_{B}$ ensures a symmetrization of the state and a lowering of the $|VV\rangle \langle VV|$ component,
leading to a higher concurrence: Fig.(3-\textbf{a}). The intensity of filtration is quantified by a parameter $0<\varepsilon\leq{1}$, connected to the attenuation of $V$ polarization. The concurrence
has a limit for asymptotic filtration ($\varepsilon \rightarrow {0}$) lower
than unity and maximal entanglement cannot be approached. This is a
cost of reversal of entanglement for distinguishable photons.
Of course, the collective protocols can be still used,
since all entangled two-qubit state are distillable to a singlet one.
Applying the filtration with the parameters $A_{A}=0.33$ and $%
A_{B}=1$, we obtain the final experimental state shown in Fig.(4-\textbf{%
III}). Such filtration brings to a theoretical concurrence of $C_{III}=0.42$ and an overall success rate of $P_{III}P_ {II}=0.17$. Hence we achieve a fidelity $F_{III}=(0.89 \pm 0.06)$ and measure a concurrence equal to $\widetilde{C}_{III}=(0.28\pm 0.02)>\widetilde{C}_{II}$
while $\widetilde{P}_{III}\widetilde{P}_{II}=(0.11\pm 0.01)$. 
The experimental results shown above represent an evident prove of the restoration protocol validity, indeed an entanglement breaking channel can be corrected to a channel
preserving relatively large amount of the entanglement.
  
\begin{figure}[b]
\centering
\includegraphics[width=0.46\textwidth]{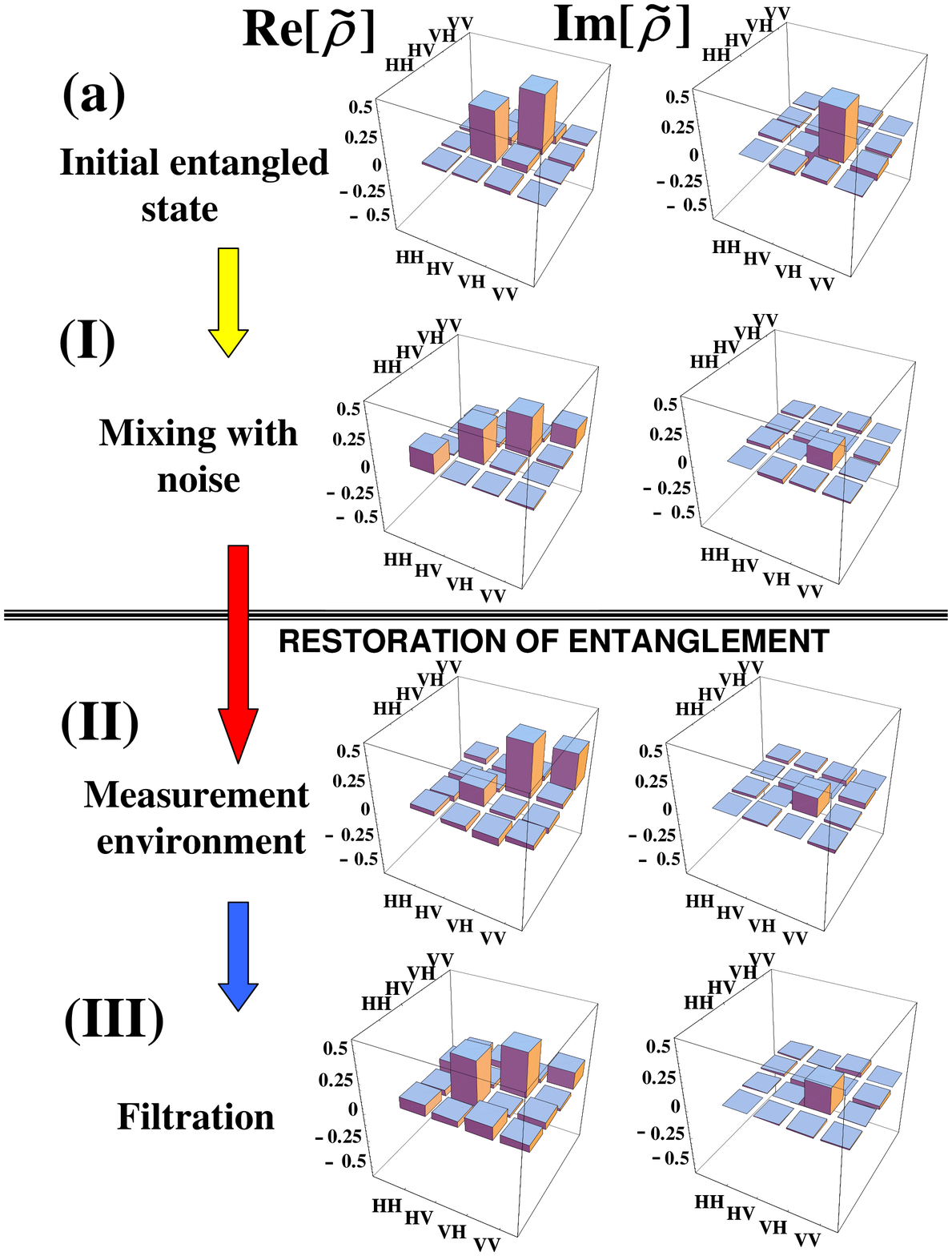}
\caption{Experimental density matrix for distinguishable photons: \textbf{a}%
) $\widetilde{\protect\rho }_{AB}^{in}$, \textbf{I}) $\widetilde{\protect%
\rho }_{AB}^{I}$, \textbf{II}) $\widetilde{\protect\rho }_{AB}^{II}$, 
\textbf{III}) $\widetilde{\protect\rho }_{AB}^{III}. $ The
fidelity with the entangled state $\left| \Psi ^{-}\right\rangle _{AB}$ is $%
\mathcal{F}(\widetilde{\rho }_{AB}^{in},\left| \Psi ^{-}\right\rangle
_{AB})=\left\langle \Psi ^{-}\right| _{AB}\widetilde{\rho }_{AB}^{in}\left|
\Psi ^{-}\right\rangle _{AB}=(0.915\pm 0.002)$ while the concurrence \cite{Woot98} reads $%
C=(0.869\pm 0.005)$. The uncertainties on the different observables have
been calculated through numerical simulations of counting fluctuations due
to the poissonian distributions. We reconstruct the two
qubit density matrix $\protect\rho _{AB}$ through the quantum state
tomography procedure \protect\cite{Jame01}. An overcomplete set of
observables is measured by adopting different polarization settings of the $%
\frac{\protect\lambda }{2}$ wp and $\frac{\protect\lambda }{4}$ wp positions.
For each tomographic setting  the measurement lasts from 5 s ($\textbf{a}$) to 30 minutes ($\textbf{III}$), the last case corresponding to about $500$ triple coincidence. Contributions due to triple accidental coincidences have been subtracted from experimental data.}
\end{figure}

\textbf{Indistinguishable noise photons}\newline
\textbf{I}) In the indistinguishable photons regime, we indicate with ${\sigma}%
^{I}_{AB}$ the density matrix after the mixing on the $BS$, which is separable for $T< 1/\sqrt{3}$
(see Table.1). The output state is found to be a Werner state, that is, a mixture of the singlet
state and the fully mixed one \cite{Wern89}.

\textbf{II}) A measurement is carried out on the
environmental mode. When the result $|H\rangle_{E}$ is obtained, ${\sigma%
}^{I}_{AB}$ evolves into ${\sigma}^{II}_{AB}$. The Werner
state is conditionally transformed into a maximally entangled state (MEMS) \cite{Vers01}.

\textbf{III}) Due to the strong unbalancement of ${\sigma}^{II}_{AB}$ a filter
is introduced either on Alice or Bob mode, depending on the value of $T$: Fig.(2). 
In the limit of asymptotic
filtration ($\varepsilon \rightarrow 0$), the concurrence reaches unity
 except for $T=1/2$ (Fig. 3-\textbf{b}). Since maximally
entangled state can be approached with an arbitrary
precision  \cite{Vers01b}, the entanglement breaking channel can
be transformed into a secure channel for key distribution.

\textbf{General Case}\newline
Let us now face up to a model which contemplates a situation close to the
experimental one. We consider
the density matrix ${\tau}_{AB}$ of the state shared between Alice and
Bob after the coupling with the environment, as a mixture arising from coupling with a partially distinguishable noise photon. The degree of indistinguishability is parametrized by the probability $p$ that the fully depolarized environmental photon is completely indistinguishable from signal: Fig.(3-\textbf{c}-\textbf{d}).

\begin{figure}[t]
\centering
\includegraphics[width=0.5\textwidth]{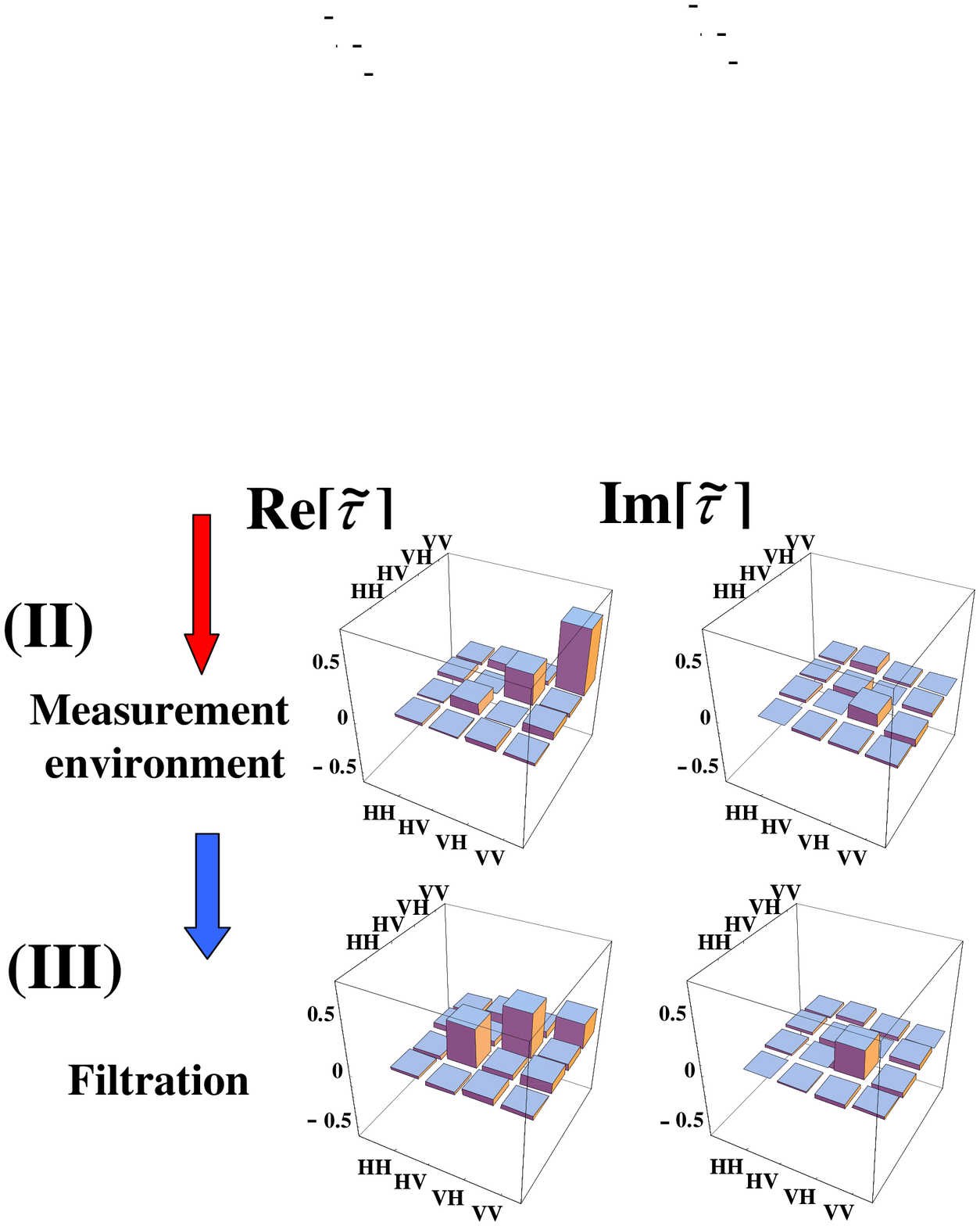}
\caption{Experimental density matrix for partially indistinguishable photons
($p=0.85$): \textbf{II}) $\widetilde{\protect%
\tau }_{AB}^{II}$, \textbf{III}) $\widetilde{\protect\tau }_{AB}^{III}$.  
Contributions due to 2-coherent environmental photon $(<10\%)$ have been estimated through triple accidental coincidences and subtracted from experimental data.}
\end{figure}

Comparing both the discussed indistinguishable
and distinguishable photons, it is evident
that it can be advantageous to induce indistinguishability.
For example, if they are distinguishable in time, then
spectral filtering can help us to make them more indistinguishable
and consequently, the entanglement can be enhanced
more by local filtering. In order to ensure a high indistinguishability between the photon of modes $%
k_{B}$ and $k_{E}$ we adopt narrow band interference filters (Fig.2). We have estimated the degree of distinguishability between the photon
belonging to mode $k_{B}$ and the one associated to $k_{E}$ as $p=(0.85\pm 0.05)$ by
realizing an Hong-Ou-Mandel interferometer \cite{Hong87} adopting a 
beamsplitter with $T=0.5$. We attribute the mismatch of $p$ with
the unit value to a different spectral profile between the coherent beam and
the fluorescence. To carry out the experiment we adopt a beamsplitter with $T=0.3$ and the optical delay has been set in the position $\Delta t=0$.
\textbf{I}) The input singlet state evolves into a noisy one represented
by $\widetilde{\tau }_{AB}^{I}$ characterized by a fidelity with theory : $%
\mathcal{F(}\widetilde{\tau }_{AB}^{I},\tau_{AB}^{I} )=(0.86\pm 0.02)$ and vanishing concurrence ($\widetilde{C}=0$).
\textbf{II}) After measuring the photon on mode $k_{E^{\prime }}$, the
density matrix $\widetilde{\tau }_{AB}^{I}$ evolves into $\widetilde{%
\tau }_{AB}^{II}$: Fig.(5-\textbf{II.}) The entanglement is restored with a
concurrence equal to $\widetilde{C}_{II}=(0.15\pm 0.03)>0$ to be compared with $C_{II}=0.22$; in this case the probability of
success reads $\widetilde{P}_{II}=(0.22\pm 0.01)$, theoretically $P_{II}=0.2$. The fidelity with the
theoretical state is $\mathcal{F(}\widetilde{\tau }_{AB}^{II},\tau
_{AB}^{II})=(0.96\pm 0.01).$ \textbf{III}) Applying the filtration with
the parameters $A_{A}=0.12$, and $A_{B}=0.30$
we obtain the state shown in Fig.(5-\textbf{III}). Hence we measure a higher
concurrence $\widetilde{C}_{III}=(0.50\pm 0.10)>\widetilde{C}_{II}$ while the expected theoretical value is $C_{III}=0.47$. The filtered state has $F(\widetilde{\tau 
}_{AB}^{III},\tau _{AB}^{III})=(0.92\pm 0.04),$ and is postselected with an overall success rate equal to $\widetilde{P}_{III}\widetilde{P}_{II}=(0.020\pm 0.002)$, where theoretically $P_{III}P_{II}=0.016$. This is a clear experimental demonstration of how an induced indistinguishability enhances the restored concurrence after an entanglement breaking channel.

In summary, we have reported the experimental demonstration of a new
protocol able to restore entanglement on entanglement breaking channel by
measuring the information leaking out into the environment. The restoration
entanglement procedure can contribute to develop a new class of
measurement-induced operations, based on cross application of single-photon
detection and feed-forward. Even for the channel preserving entanglement, it can be
particularly advantageous to detect environmental state
outgoing from the coupling, because the entanglement
can be increased simply by local filtering on single copy instead of more demanding collective
distillation protocol. The present scheme can be generalized to restore entanglement after consecutive interactions with different sources of noise by properly measuring the outgoing environment. Moreover, a direct application can be envisaged in quantum
computation with matter devices, where decoherence can be attributed to
coupling of signal qubits with other degrees of freedom of the information
carrier. In this framework, a measurement on this available environment can
be adopted to retrieve entanglement. On the fundamental side the present
experiment provide new elements to overcome decoherence effects, a subject
of renewed interest in the last year \cite{Alme07} \cite{Konr08}.

F.D.M., E.N. and F.S. acknowledge support by the PRIN 2005 of MIUR and Progetto Innesco 2006
(CNISM). R.F. and M.G. have been supported by MSM 6198959213 and LC 06007 of Czech Ministry of Education, AvH Foundation and Grant 202/08/0224 of GACR.

\end{document}